\documentstyle[12pt]{article} 
\newcommand{\be}{\begin{equation}}
\newcommand{\ee}{\end{equation}} 
\newcommand{\ba}{\begin{eqnarray}}
\newcommand{\ea}{\end{eqnarray}} 
\begin{document}
%%%%%%%%%%%%%%%%%%%%%%%%%%%%%%%%%%%%%%%%%%%%%%%%%%%%%%%%%%%%%%%%%%%%%
\title {Decoupling Transformations\\ in Path Integral Bosonization} 
\author{C.D. Fosco$^a$\thanks{Investigador Conicet, Argentina}
\\and\\
J.C. Le Guillou$^b$\thanks{Also at {\it Universit\'e de Savoie} 
and at {\it Institut Universitaire de France}}\\ 
{\normalsize\it $^a$Centro At\'omico Bariloche, 
Comisi\'on Nacional de Energ\'{\i}a At\'omica}\\
{\normalsize\it 8400 Bariloche, Argentina}\\~\\
{\normalsize\it $^b$Laboratoire  d'Annecy-le-Vieux
 de Physique Th\'eorique LAPTH}
\thanks{URA 1436 du CNRS associ\'ee \`a l'Universit\'e de Savoie}\\
{\normalsize\it LAPP, B.P. 110, F-74941 Annecy-le-Vieux Cedex, France}} 
\date{} 
\maketitle 
\vspace{-5.1in} 
\hfill\vbox{\hbox{\it LAPTH-685/98}} 
\vspace{4.5 in} 
\begin{abstract} 
We construct transformations that decouple fermionic fields in 
interaction with a gauge field, in the path integral representation of 
the generating functional.  Those transformations express the original 
fermionic fields in terms of non-interacting ones, through non-local 
functionals depending on the gauge field.  This procedure, holding 
true in any number of spacetime dimensions both in the Abelian and 
non-Abelian cases, is then applied to the path integral bosonization 
of the Thirring model in $3$ dimensions.  Knowledge of the decoupling 
transformations allows us, contrarily to previous bosonizations, to 
obtain the bosonization with an explicit expression of the fermion 
fields in terms of bosonic ones and free fermionic fields.  We also 
explain the relation between our technique, in the two dimensional 
case, and the usual decoupling in $2$ dimensions.
\end{abstract} 
\newpage
%===================================================================
\section{Introduction} 
Any physical system can be studied by using different sets of variables
which yield, of course, {\em mathematically\/} different descriptions.
Different choices of variables must however be {\em physically\/}
equivalent, in the sense that they should describe one and the same
system. An extreme manifestation of this property appears in models
which can be defined in terms of either fermionic or bosonic variables.
The equivalence between these two formulations is made explicit by the
so called `bosonization rules', that map fermionic into bosonic
variables. 

Bosonization turns out to be a very useful tool indeed in order to 
understand and in some cases even to solve some non-trivial 
interacting Quantum Field Theory models in two spacetime dimensions.  
It is an interesting fact that there is no theoretical obstacle to the 
extension of this procedure to higher dimensions.  Indeed, there has 
recently been some progress in the application, although in an 
approximated form, of a path integral bosonization procedure to 
theories in more than two spacetime dimensions~\cite{BQ}-\cite{LGMNS}, 
dealing with both the Abelian and the non-Abelian cases.  An important 
difference between these works and two dimensional 
bosonization~\cite{fur}-\cite{joli} is that, in the latter case, the 
fermionic determinant can be evaluated by performing a decoupling 
transformation of the fermions~\cite{gam3}-\cite{poly}.  The redefined 
fermionic fields (in the massless case) are free, and the effects of 
the interaction manifest themselves through the existence of an 
anomalous Jacobian which depends on the gauge field.  We remark that 
no such a thing had yet been suggested for the higher dimensional 
case; the present work deals with such an extension.  In this paper we 
shall complement the usual path integral bosonization technique by 
showing how to redefine the fermionic fields in the path integral, in 
order to decouple them from the gauge field.  This step, we believe, 
fills a gap in this approach to bosonization in higher dimensions, 
rendering the whole process entirely analogous to its well known two 
dimensional counterpart.  The main point of our work is that it gives, 
contrarily to previous bosonizations in 3 dimensions, the bosonization 
with an explicit expression for the fermionic fields in terms of 
bosonic ones and free fermionic fields.  This procedure is moreover 
generalizable to higher dimensions.

Let us briefly review, for the Abelian case in $3$ dimensions, the usual
approach to bosonization, which starts from the definition of the
generating functional ${\cal Z} (J)$, 
$$
{\cal Z} (J) \;=\; \int {\cal D} {\bar \Psi} {\cal D} \Psi \, 
\exp \left[- \int d^3 x \, {\bar \Psi} (\not \! \partial
+ \not \! J + M ) \Psi \right] 
$$
\be
=\;\det \left[\not \! \partial + \not \! J + M   \right]
\;=\; \exp \left[ - W(J) \right]
\label{zfer}
\ee
where $W(J)$ is the generating functional of Euclidean connected correlation 
functions of fermionic currents.
The bosonized action $S_{bos} (A)$ is then given by a generalised functional
Fourier transform of $Z(b)= \exp [-W(b)]$:
\be
\exp[-S_{bos}(A)]\;=\; \int {\cal D} b_\mu \, \exp \left[ - W(b) +i
\int d^3 x \epsilon_{\mu\nu\lambda} b_\mu \partial_\nu A_\lambda
\right]\;,
\label{sbos}
\ee
and the bosonization rule for the fermionic bilinear ${\bar \Psi}
\gamma_\mu \Psi$ is
\be
{\bar \Psi}(x)\gamma_\mu \Psi(x) \;\to \; i \epsilon_{\mu\nu\lambda} 
\partial_\nu A_\lambda(x) \;.
\ee
 
The fact that the functional $W(J)$ proceeds from fermionic
matter field seems to be here a matter of no relevance, the crucial
property invoked being gauge invariance of the fermionic determinant
instead. This is quite different to what happens in two dimensional path
integral bosonization, where the fermionic fields are redefined in terms
of new, decoupled fermions, through transformation functions that depend
on the bosonic gauge fields. In this paper, we give explicit
transformation functions that achieve the same goal as in the two
dimensional case.

A difficulty renders the bosonization procedure in more than two
spacetime dimensions non-exact. It is our inability to compute exactly a
fermionic determinant in those cases. This allows us to determine only
approximated bosonized actions, which usually result from some perturbative
or low-momen-tum expansion. We shall see that the same can be said about
the decoupling transformations, one can consistently decouple the fermions
up to some order in the relevant coupling constant of the model being
considered.

The structure of this paper is as follows: In section 2 we explain the
mechanism of decoupling on a general footing, discussing both Abelian
and non-Abelian cases, in the massless and massive situations. The
application of the general results to the path-integral bosonization of
the three dimensional Thirring model is studied in section 3, and in
section 4 we explain the differences and similarities between our
approach, when restricted to the two dimensional case, and the one which
makes use of the anomalous (Fujikawa) Jacobian method.

\newpage
\section{Decoupling the fermions by a change of variables}
\subsection{Massless case}
The generating functional ${\cal Z}(A)$ for massless fermionic fields in
the presence of an external gauge field $A$ in a $D$-dimensional
Minkowski spacetime~\footnote{We shall later on explain the analogous 
procedure for the Euclidean spacetime case.} is defined by
\be
{\cal Z}(A) \;=\;\int {\cal D}{\bar \Psi}\,{\cal D}\Psi \; 
e^{i S_F({\bar \Psi},\Psi;A)} 
\label{dfza}
\ee
where 
$$
S_F({\bar \Psi},\Psi;A)\;=\; \int d^Dx \;
{\bar \Psi}(x) \, i\not \!\! D \,\Psi (x)
$$
\be
\not \!\! D\,=\,\gamma_\mu D_\mu\;\;\;\;\;\;
D_\mu\,=\,\partial_\mu + e A_\mu
\ee
and 
\be
A \;=\;-A^\dagger\;\equiv\; 
\left\{ 
\begin{array}{c} 
i {\cal A}_\mu \;\;{\rm in \,the\, Abelian\, case}\\
{\cal A}_\mu^a \tau_a \;\; {\rm in\, the\, non\, Abelian\, case}
\end{array}
\right.\;,
\ee
where $\tau_a$ are (anti-hermitian) generators of the Lie algebra of the
non-Abelian gauge group, and both ${\cal A}_\mu$ and ${\cal A}_\mu^a$
are real. 
Dirac's $\gamma$-matrices are assumed to satisfy
\be
\{ \gamma_\mu , \gamma_\nu \} \;=\; 2 g_{\mu\nu} \;\;\quad 
\gamma_0^\dagger = \gamma_0 \;\;\quad \gamma_j^\dagger = - \gamma_j
\;.
\ee

We shall show that, by a change of variables, one can decouple 
the fermions from the external field $A$. 
Then we redefine the fermionic fields $\Psi (x)$, ${\bar \Psi}(x)$ in
terms of new ones $\chi (x)$, ${\bar \chi}(x)$, as follows 
\ba
\Psi (x) &=& {\cal F}(e{\not \! \partial}^{-1} \not \!\! A)\,\chi (x)
\nonumber\\
{\bar \Psi}(x) &=& {\bar \chi}(x) 
{\bar {\cal F}}(e {\not \! \partial}^{-1} \not \!\! A)
\label{deft1}
\ea
where 
\be
{\bar {\cal F}} (e {\not \! \partial}^{-1} \not \!\! A)\,=\,
\gamma_0 \left[{\cal F}(e{\not \! \partial}^{-1} \not \!\!
A)\right]^\dagger \gamma_0
\ee 
and ${\cal F}$ is a function which, we shall assume, is defined by a
power series expansion in $e$
\be
{\cal F}(e{\not \! \partial}^{-1} \not \!\! A)\;=\; 1 + \sum_{n=1}^\infty
\, \alpha_n\, e^n\, ({\not \! \partial}^{-1}\not \!\! A)^n \;,
\label{pwsr}
\ee
with real coefficients $\alpha_n$. The fact that we shall understand the
transformation (\ref{deft1}) as given by the series (\ref{pwsr}) is the
reason why we call the whole technique `perturbative'. 

It is important to note that the operator ${\not \! \partial}^{-1}$
introduced in (\ref{deft1}) must be understood as acting on both 
$\not\!\!A$ {\em and\/} the fermionic field $\chi$, not on 
$\not\!\!A$ alone. Namely,
\be
{\cal F}(e{\not \! \partial}^{-1} \not \!\! A)
\neq
{\cal F}(e\,({\not \! \partial}^{-1} \not \!\! A)\,)\;,
\ee
where the outer parenthesis denotes functional dependence, while the
inner one delimites the action of the operator ${\not \!\partial}^{-1}$.
This fact and the reality of the $\alpha_n$'s implies that the transformation 
formula for ${\bar \Psi}(x)$ may also be written more explicitly as
\be
{\bar \Psi}(x) \;=\; {\bar \chi}(x) \, {\cal F}(e\not \!\! A 
{\not \!\partial}^{-1})
\ee
and this property turns out to be crucial in the decoupling. We then
see that the new fields will have a free action as long as the equation
\be
\int d^D x\; {\bar \chi}\, {\cal F}(e\not \!\! A {\not \! \partial}^{-1})
\, i\not \!\! D\, {\cal F}(e{\not \! \partial}^{-1} \not \!\! A)\, \chi(x)
\;=\; \int d^D x\; {\bar \chi}\, i\not \!\partial\, \chi (x) 
\label{cnd1}
\ee
may be satisfied.
This implies 
\be
{\cal F}(e\not \!\! A {\not \! \partial}^{-1})\;
\not \!\! D
\;{\cal F}(e{\not \! \partial}^{-1} \not \!\! A)
\;=\; \not \!\partial
\ee
which, by using the power series expansion (\ref{pwsr}), yields a set of
relations for the unknown coefficients $\alpha_n$. They may be written
as
$$
0\;=\;e (1 + 2 \alpha_1) \not \!\! A + \sum_{n=1}^\infty 2 (\alpha_n +
\alpha_{n+1}) e^{n+1} T^{(n)} + 
$$
\be
\sum_{n=1}^\infty (\sum_{m=1}^n \alpha_{n+1-m} \alpha_m ) e^{n+1} T^{(n)}
\,+\,\sum_{n=2}^\infty (\sum_{m=1}^{n-1} \alpha_{n-m} \alpha_m ) e^{n+1} T^{(n)}
\label{rels}
\ee
where 
\be
T^{(n)} \;=\; \not \!\! A ({\not \! \partial}^{-1} \not \!\! A)^n
\;=\; (\not \!\! A {\not \! \partial}^{-1})^n \not \!\! A 
\;=\; \not \!\! A {\not \! \partial}^{-1} \not \!\! A \cdots
{\not \! \partial}^{-1} \not \!\! A \;
\ee
namely, the operator $T^{(n)}$ consists of a product of alternating
factors ${\not \! \partial}^{-1}$ and $\not \!\! A$, starting and ending
with an $\not \!\! A$, and containing $n$ factors ${\not \!
\partial}^{-1}$ altogether. The crucial property that makes the
decoupling possible is that any redefinition like (\ref{deft1}), (even
with arbitrary coefficients $\alpha_n$), modifies the action by adding a
series of terms, which are always proportional to one of the
$T^{(n)}$'s. 

Equation (\ref{rels}) implies the recurrence relations
\be
\alpha_{n+1} \;=\; -\frac{3}{4} \alpha_n \,-\,\frac{1}{2}
\sum_{k=1}^{n-1} (\alpha_{k+1}+\alpha_k) \alpha_{n-k} \;\;\forall n>1
\label{recrel}
\ee 
plus the initial conditions
\be
\alpha_1 = - \frac{1}{2} \;\;\; \alpha_2 = \frac{3}{8} \;.
\label{inc}
\ee
The solution to the recurrence relations (\ref{recrel}), with conditions
(\ref{inc}), is
\be
\alpha_n \;=\; \left(\begin{array}{c}-\frac{1}{2}\\ n 
\end{array}\right)
\ee
where
\be
\left(\begin{array}{c}-\frac{1}{2}\\ n 
\end{array}\right)\;=\; \frac{(-\frac{1}{2})(-\frac{1}{2}-1) \cdots 
(-\frac{1}{2}-n+1)}{n!} \;.
\ee
Thus we conclude that ${\cal F}(x)$ can be thought of as the power
series expansion of the function $(1 + x)^{-\frac{1}{2}}$ around $x=0$,
with a radius of convergence equal to $1$,
\be
{\cal F}(x) \;=\; \sum_{n=0}^\infty 
\left(\begin{array}{c}-\frac{1}{2}\\ n \end{array}\right)\,x^n\;=\;
(1 + x)^{-\frac{1}{2}} \;, \;\; |x| < 1 \;.
\label{pwr}
\ee
Condition $|x|<1$ is equivalent to saying that we are in the
perturbative regime. Whenever we write, henceforward, ${\cal F}(x) \;=\;
(1 + x)^{-\frac{1}{2}}$, we shall have in mind its power series
expansion as given by (\ref{pwr}).

When performing the transformations
\ba
\Psi (x) &=& (1 + e{\not \! \partial}^{-1} \not \!\! A)^{-\frac{1}{2}}\,
\chi (x)\nonumber\\
{\bar \Psi}(x) &=& {\bar \chi}(x)(1 + 
e\not \!\! A {\not \! \partial}^{-1} )^{-\frac{1}{2}} \;,
\label{deft2}
\ea
we can then affirm that the new fermionic action shall be free, but, on
the other hand, there will appear a Jacobian $J(A)$, due to the change
of (Grassmann) variables, 
$$ 
{\cal Z}(A) \;=\;\int {\cal D}{\bar \Psi}\,{\cal D}\Psi \; 
e^{i S_F({\bar \Psi},\Psi;A)} 
$$
\be
=\; \int {\cal D}{\bar \chi}\,{\cal D}\chi \; J(A) \;
e^{i S_F({\bar \chi},\chi;0)} 
\label{dex}
\ee
where,
\be
S_F({\bar \chi},\chi;0)\;=\; \int d^Dx \,
{\bar \chi}(x) \, i\not \! \partial\, \chi (x)
\ee
and
$$
J(A) \;=\; J_\chi (A) \, J_{\bar{\chi}} (A) \;,
$$
$$
J_\chi (A) \;=\; \left\{\det[(1+e{\not\!\partial}^{-1}\not 
\!\!A)^{-\frac{1}{2}}]\right\}^{-1}
$$
\be
J_{\bar{\chi}} (A)\;=\; \left\{\det[(1+e\not \!\!A 
{\not\!\partial}^{-1})^{-\frac{1}{2}}]\right\}^{-1} \;.
\ee
We remark that these Jacobians are different from one, but nevertheless
not anomalous, in the sense that, as the field transformations are not
unitary, they shouldn't be equal to one.

From the two self-evident relations
\ba
\det \not \!\! D &=& \det \not \! \partial \, 
\det(1+e{\not\!\partial}^{-1}\not \!\!A) \nonumber\\
\det \not \!\! D &=&\det(1+e\not \!\!A {\not\!\partial}^{-1})\,
\det \not \! \partial
\ea
we deduce
\be
J(A) \;=\; \frac{\det (\not \!\! D)}{\det (\not \!\partial)} \;.
\label{jac1}
\ee
Recalling (\ref{dex}), we see that, as it should be, multiplying the
Jacobian by the free determinant we reconstruct the determinant of
$\not \!\! D$. 

The use of transformations (\ref{deft2}) would, in principle, allow us
to decouple the fermions completely from the external field $A$. 
This, however, is not possible to use in the applications, since it would 
require the evaluation of the Jacobian (\ref{jac1}) exactly, something not possible in more than $1+1$
dimensions. We may, however, still use a finite order approximation to
(\ref{deft2}), in order to achieve {\em partial\/} decoupling. For
example, we can perform a transformation that corresponds to the
truncation to second order in $e$ of (\ref{deft2}). This will lead to an
action for the new fermions containing interaction terms of order three
or higher; namely,
\ba
\Psi (x) &=& \left( 1-\frac{1}{2} e {\not \! \partial}^{-1} \not \!\! A
\,+\, \frac{3}{8} e^2 {\not \! \partial}^{-1} \not \!\! A
{\not \! \partial}^{-1} \not \!\! A \right)\, \chi (x) \nonumber\\
{\bar \Psi}(x) &=& {\bar \chi}(x)\left( 1-\frac{1}{2} e 
 \not \!\! A{\not \! \partial}^{-1}
\,+\, \frac{3}{8} e^2  \not \!\! A{\not \! \partial}^{-1}
\not \!\! A {\not \! \partial}^{-1}\right)
\ea
and
\be
S_F \to \int d^D x \; {\bar \chi} ( i \not \! \partial + {\cal O} (e^3)
) \chi \;.
\ee

The use of the decoupling transformations in the {\em Euclidean\/}
spacetime case could of course be implemented by first rotating to 
Minkowski spacetime, then performing the transformations (\ref{deft2}),
and at the end coming back to Euclidean. The new fermions will
of course remain decoupled after the rotation. 
If one wanted to implement the transformations directly in Euclidean spacetime, 
one should face the problem that the Dirac matrices have different
hermiticity properties than in Minkowski spacetime, and that spoils the
transformation rule for the adjoint spinor, since under the 
transformation
\be
\Psi (x) \;=\; {\cal F}(e{\not \! \partial}^{-1} \not \!\! A)\,\chi (x)
\ee
it is still true that 
\be
{\bar \Psi}(x) \;=\; {\bar \chi}(x) \,{\bar {\cal F}} 
(e {\not \! \partial}^{-1} \not \!\! A)
\ee
but
\be
\gamma_0 \left[{\cal F}(e{\not \! \partial}^{-1} \not \!\!
A)\right]^\dagger \gamma_0 \;=\; {\cal F}(e\not \!\! {\tilde A} 
{\not \!{\tilde \partial}}^{-1})
\;\neq\; 
{\cal F}(e\not \!\! {\tilde A} {\not \!{\tilde \partial}}^{-1})
\ee
where, for any Euclidean vector $v_\mu$, we denote 
${\tilde v}_\mu = (-v_1,-v_2,v_3)$. 

Although not necessary in principle, one can, however, still give an 
explicit decoupling transformation in the Euclidean case. To that end, 
we employ the trick used in \cite{zinn}, where it was introduced in a 
slightly different context. 
That trick amounts to make in the Euclidean path integral representing 
the fermionic determinant
\be
{\cal Z}(A) \;=\;\int {\cal D}{\bar \Psi}\,{\cal D}\Psi \; 
e^{- \int d^Dx {\bar \Psi}(x) \,\not \! D \,\Psi (x)} 
\;=\; \det \not \!\! D \;,
\label{dfzeu}
\ee
the change ${\bar \Psi} (x) \to {\bar \Psi} (x) \gamma_3$, which of
course does not change the value of the determinant~\footnote{One
could of course have just said that one wanted to evaluate the determinant
of $\gamma_3 \not \!\! D$.}.
Then one see that the fermionic action is
\be
S_F \;=\; \int d^Dx {\bar \Psi}(x) \gamma_3 \not \!\! D \,\Psi (x)
\ee
and, under the decoupling transformations
\ba
\Psi (x) &=& (1 + e{\not \! \partial}^{-1} \not \!\! A)^{-\frac{1}{2}}\,
\chi (x)\nonumber\\
{\bar \Psi}(x) &=& {\bar \chi}(x) \gamma_3 (1 + 
e\not \!\! A {\not \! \partial}^{-1} )^{-\frac{1}{2}}
\gamma_3 \;,
\label{defte}
\ea
the new action is 
\be
S_F \;\to\; \int d^Dx {\bar \chi}(x) \gamma_3 \not \!\partial \,\chi (x)
\ee
as can be easily checked. Note that the presence of $\gamma_3$ in the
new action is as irrelevant as in the old one. To avoid writing the
$\gamma_3$, when we use the Euclidean formalism in section 3, 
we work with $\Psi^\dagger$ instead of ${\bar \Psi}$, 
\be
S_F \;=\; \int d^Dx \Psi^\dagger (x) \not \!\! D \,\Psi (x)
\ee
and give the transformations in terms of those fields. The
procedure is, of couse, equivalent.

\subsection{Massive case}
The massive case admits, within our decoupling approach, two different
treatments. Beginning from the massive fermionic action
\be
S_F({\bar \Psi},\Psi;A)\;=\;\int d^Dx\;
{\bar \Psi}(x)\, (i \not \!\! D - M)\, \Psi (x) \;,
\label{masac}
\ee
one possibility is to perform the analogous of the redefinitions 
(\ref{deft2}) but including the mass term into the inverse of the free Dirac 
operator, i.e.,
\ba
\Psi (x) &=& [1 + e (\not \! \partial + i M)^{-1} 
\not \!\! A]^{-\frac{1}{2}}\,\chi (x)\nonumber\\
{\bar \Psi}(x) &=& {\bar \chi}(x)\,[1 + 
e\not \!\! A (\not \! \partial + i M)^{-1} ]^{-\frac{1}{2}} \;.
\label{deftt}
\ea
This way of transforming the fermionic fields leads, after some algebra
that is essentially equal to the one of the massless case, to the
decoupled fermionic fields ${\bar \chi}$ and $\chi$, with a massive free
action
\be
S_F({\bar \chi},\chi)\;=\; \int d^Dx 
{\bar \chi}(x)\,(i \not \! \partial - M) \,\chi (x) \;,
\ee
and to the Jacobian
\be
J(A) \;=\; \frac{\det (\not \!\! D + i M)}{\det 
(\not \!\partial + i M)} \;;
\label{jac2}
\ee
this way of decoupling, applied in its second order approximated
version, is the one we use in the next section to study the Thirring
model in $2+1$ dimensions.

There is, however, another way of transforming the fermions, which is in
fact more similar to the one used in two dimensions. It consists in
using the transformations (\ref{deft2}), corresponding to the massless
case, for the massive action. The new fermions are no longer decoupled,
but all the interactions are proportional to the mass, and the new
action is a suitable starting point for performing an expansion in
powers of the mass, as it happens to be the case in two dimensions 
$$
S_F({\bar \chi},\chi; A)\;=\; \int d^Dx 
{\bar \chi}(x) i\not \! \partial \chi (x) 
$$
\be
- M \int d^D x {\bar \chi} (x)
\left\{
[1 + e(\not \! \partial + i M)^{-1} \not \!\! A]
[1 + e\not \!\! A (\not \!\! \partial + i M)^{-1} ]
\right\}^{-\frac{1}{2}}
\chi (x) \;.
\ee

In Euclidean spacetime, as for the massless case, the continuation to
Minkowski prior to decoupling is a possible path. However, one can also
work directly in Euclidean spacetime as explained for the massless case.
One then uses the $\gamma_3$ trick. If the first version of decoupling
(\ref{deftt}) is used, due to the presence of the mass term, we must
also make a "smooth continuation" (in the sense of \cite{ball}) to
imaginary masses before decoupling, and then rotate back to real
values. In perturbation theory, there is no obstruction to that.
Then the transformations in terms of $\Psi$ and $\Psi^\dagger$ read
\ba
\Psi (x) &=& [1 + e (\not \! \partial + M)^{-1} 
\not \!\! A]^{-\frac{1}{2}}\,\chi (x)\nonumber\\
\Psi^\dagger (x) &=& \chi^\dagger (x)\,[1 + 
e\not \!\! A (\not \! \partial + M)^{-1} ]^{-\frac{1}{2}} \;.
\label{deft3}
\ea
where $M$ is regarded as imaginary.

We end up this section by making a consistency check, valid for both the
massive and the massless cases. Namely, that the expectation value of the 
fermionic current ${\bar\Psi} \gamma_\mu \Psi$, when the fermions are written in terms of the
$\chi$ and ${\bar \chi}$, yields the proper result. 
For the sake of concreteness, we deal the case of the massive action 
(\ref{masac}). The expectation value of that current, when evaluated in terms of the
decoupled fermions is 
$$
\langle {\bar \Psi} (x) \gamma_\mu \Psi (x) \rangle \;=\; 
$$
\be
\langle {\bar \chi}(x) 
[1 + e\not \!\! A (\not \! \partial + i M)^{-1} ]^{-\frac{1}{2}}
\gamma_\mu 
[1 + e (\not \! \partial + i M)^{-1} \not \!\! A]^{-\frac{1}{2}}\,\chi (x)
\rangle \;,
\ee
or
$$
\langle {\bar\Psi} (x) \gamma_\mu \Psi (x) \rangle \;=\; 
$$
$$
-{\rm Tr}\int d^D y d^D z \left\{  
\langle y|[1 + e\not \!\! A (\not \! \partial + i M)^{-1} ]^{-\frac{1}{2}}
|x \rangle \gamma_\mu 
\right.
$$
\be
\left.
\langle x|
[1 + e (\not \! \partial + i M)^{-1} \not \!\! A]^{-\frac{1}{2}}
|z\rangle \,\langle\chi (z) {\bar \chi}(y)\rangle \right\}
\ee
where the usual notation $\langle x | \cdots |y \rangle$  denotes kernels 
of functional operators in coordinate space.

Since the new fermions are free,
\be
\langle\chi (z){\bar \chi}(y)\rangle \;=\; \langle z | 
(\not \! \partial + i M)^{-1} |y \rangle \;.
\ee
Using this, and after some straightforward algebra, we obtain 
\be
\langle {\bar\Psi} (x) \gamma_\mu \Psi (x) \rangle \;=\; 
-{\rm Tr}\left[ \gamma_\mu 
\langle x|(\not \!\! \partial +i e \not \!\! A + i M)^{-1}
|x \rangle \right]
\label{exte}
\ee
which is the proper expression for the expectation value of the
current in an external field. Also, (\ref{exte}) can of course be
written in terms of the fermionic determinant:
\be
\langle {\bar\Psi} (x) \gamma_\mu \Psi (x) \rangle \;=\; 
\frac{\delta}{e \delta A_\mu (x)} \ln \det (i\not \!\! D - M) \;.
\label{pet2}
\ee

\newpage
\section{The Thirring model in $2+1$ dimensions}
We shall here follow the same strategy as in \cite{joli} for the $1+1$
dimensional Thirring model, but with the necessary changes due to the
fact that we are now in $2+1$ dimensions, and the decoupling is,
in consequence, non-exact.

We define the generating functional of current correlation functions for
the Thirring model in $2+1$ Euclidean dimensions by
\be
{\cal Z}(J) \;=\; \int {\cal D}\Psi^\dagger \, {\cal D} \Psi \;
\exp \left\{- \int d^3 x [ \Psi^\dagger (\not \! \partial + \not \!\! J
+ M) \Psi + 
\frac{1}{2} g (\Psi^\dagger\gamma_\mu \Psi )^2]
\right\} 
\ee
where $g$ is the coupling constant and $J_\mu$ is a source introduced in
order to generate fermionic current correlation functions.
In order to render the fermionic action quadratic, we introduce an
auxiliary vector field $A_\mu$, such that the generating functional now
reads
$$
{\cal Z}(J) \;=\; \int {\cal D}\Psi^\dagger\,{\cal D}\Psi \;{\cal D}A_\mu 
$$
\be
\times \exp \left\{- \int d^3 x [\Psi^\dagger (\not \! \partial + 
i \sqrt{g}\not \!\! A + \not \!\! J + M)\Psi +\frac{1}{2} (A_\mu)^2 ]
\right\} \;.
\ee
Making now a shift in the vector field $A_\mu\,\to \,A'_\mu=i\sqrt{g}A_\mu+
J_\mu$,
\be
{\cal Z}(J)=\int {\cal D}\Psi^\dagger {\cal D}\Psi
{\cal D}A'_\mu \exp \left\{- \int d^3 x [ 
\Psi^\dagger (\not \! \partial + \not \!\! A' + M) 
\Psi - \frac{1}{2 g} (A'-J)^2]\right\} \,.
\ee
Now we proceed to make a decomposition of the field $A'$ into the curl
of a vector field $\Phi_\mu$, plus the gradient of a scalar field
$\varphi$,
\be
A'_\mu \;=\; i ( \epsilon_{\mu\nu\lambda} \partial_\nu \Phi_\lambda
+ \partial_\mu \varphi ) \;.
\label{dcmp}
\ee
We note that we can shift the vector field $\Phi_\mu$ by a gradient
without affecting the configuration of $A'_\mu$. This freedom is due to
the fact that $A'_\mu$ has three components, while on the right hand side
of (\ref{dcmp}) there appear to be four (three from $\Phi_\mu$ plus one
from $\varphi$). This apparent contradiction is solved by the above
mentioned `gauge invariance' under transformations of $\Phi_\mu$, which
allows we to impose a gauge fixing condition on $\Phi_\mu$, in order
to leave only two components. We choose the Lorentz condition
\be
\partial \cdot \Phi \;=\; 0 \;,
\ee
which shall, of course, be included in the generating functional.
We then get an expression for ${\cal Z}(J)$ as an integral over the fermions
$\Psi^\dagger, \Psi$ and the bosonic fields $\Phi_\mu$, $\varphi$
$$
{\cal Z}(J) \;=\; \int {\cal D}\Psi^\dagger \,{\cal D} \Psi \;
{\cal D}\Phi_\mu {\cal D}\varphi \, \delta(\partial \cdot \Phi)
$$
$$
\exp \left\{-\int d^3 x [\Psi^\dagger(\not \! \partial + i \gamma_\mu 
\epsilon_{\mu\nu\lambda}\partial_\nu \Phi_\lambda + 
i \not \! \partial \varphi + M) \Psi \right.
$$
\be
\left. -\frac{1}{2 g} (i\epsilon_{\mu\nu\lambda}\partial_\nu \Phi_\lambda 
+ i\partial_\mu \varphi - J_\mu)^2] 
\right\} \;.
\ee

By a redefinition of the fermions, 
\ba
\Psi (x) &=& e^{-i\varphi (x)} \; \Psi'(x) \nonumber\\
\Psi^\dagger (x) &=& \Psi'^\dagger(x) \; e^{i\varphi (x)} 
\ea
we entirely decouple them from the scalar 
field $\varphi$:
$$
{\cal Z}(J) = \int {\cal D}\Psi'^\dagger \; {\cal D} \Psi' \;
{\cal D}\Phi_\mu \; {\cal D}\varphi \;\;\delta(\partial \cdot \Phi)
$$
$$
\exp \left\{-\int d^3 x [\Psi'^\dagger(\not \! \partial + i \gamma_\mu 
\epsilon_{\mu\nu\lambda}\partial_\nu \Phi_\lambda + M) \Psi' \right.
$$
\be 
\left. +\frac{1}{2 g} \Phi_\mu (-\partial^2 \delta_{\mu\nu} + 
\partial_\mu \partial_\nu) \Phi_\nu + \frac{1}{2 g} 
(\partial_\mu \varphi)^2 - \frac{1}{2 g} J^2 + \frac{i}{g} J_\mu 
(\partial_\mu \varphi + \epsilon_{\mu\nu\lambda}\partial_\nu \Phi_\lambda)] 
\right\} \;.
\ee

At this point we make use of the results of the previous section, 
to perform a new redefinition of the fermions, corresponding to 
(\ref{deft3}), this time to decouple them from $\Phi_\mu$
\ba
\Psi'(x) &=& [1+(\not\!\partial+M)^{-1}(\not\!\partial 
\not\!\!\Phi) ]^{-\frac{1}{2}} \chi(x) \nonumber\\
\Psi'^\dagger(x) &=& \chi^\dagger(x) [1+(\not\!\partial \not\!\!\Phi)
(\not\!\partial+M)^{-1}]^{-\frac{1}{2}} \;,
\label{defte1}
\ea
where we have used
\be
 i \gamma_\mu \epsilon_{\mu\nu\lambda}\partial_\nu \Phi_\lambda 
 \,=\, (\not\!\!\partial \not\!\!\Phi)  
\ee
which holds as a consequence of the Dirac algebra in $3$ dimensions 
when the Lorentz condition for $\Phi$ is used. 
Transformation (\ref{defte1}) introduces a $\Phi$ dependent Jacobian,
$$
{\cal Z}(J) \;=\; \int {\cal D}\chi^\dagger \,{\cal D} \chi \;
{\cal D}\Phi_\mu {\cal D}\varphi \,J(\Phi)\delta(\partial \cdot \Phi)
$$
$$
\exp \left\{-\int d^3 x [\chi^\dagger(\not \! \partial + M) \chi 
\,+\,\frac{1}{2 g} \Phi_\mu (-\partial^2 \delta_{\mu\nu} + 
\partial_\mu \partial_\nu) \Phi_\nu 
\right.
$$
\be 
\left. 
+ \frac{1}{2 g} 
(\partial_\mu \varphi)^2 - \frac{1}{2 g} J^2 + \frac{i}{g} J_\mu 
(\partial_\mu \varphi + \epsilon_{\mu\nu\lambda}\partial_\nu \Phi_\lambda)] 
\right\} 
\ee
where
\be
J(\Phi)\;=\;\det
[1+(\not\!\partial+M)^{-1}(\not\!\partial\not\!\!\Phi)]\;.
\ee
As already explained, it is not possible to evaluate the Jacobian
$J(\Phi)$ exactly. We shall instead use the quadratic approximation
of \cite{BFO}, which yields for this object
\be
J(\Phi) \;=\; \exp [- W (\Phi) ] 
\ee
with
\be
W(\Phi)\;=\; \int d^3x\, 
\Phi_\mu 
[C^+ {\cal P}_+ + C^- {\cal P}_- ]_{\mu\nu}
\Phi_\nu 
\ee
where the $C_\pm$ are scalar operator functions \cite{BFO}
and the ${\cal P}_\pm$ are two of the three orthogonal projectors
used in the calculation:
\be
[{\cal P}_\pm]_{\mu\nu} \;=\; \frac{1}{2}(\delta_{\mu\nu}-
\frac{\partial_\mu\partial_\nu}{\partial^2} \pm
\frac{i}{\sqrt{-\partial^2}} \epsilon_{\mu\lambda\nu} 
\partial_\lambda)
\;\;\;\;
Q_{\mu\nu} \;=\; \frac{\partial_\mu\partial_\nu}{\partial^2} \;.
\ee
We shall also need the transverse projector
\be
{\cal P} \;=\; {\cal P}_+ \,+\, {\cal P}_- \;=\; \delta_{\mu\nu} 
- \frac{\partial_\mu\partial_\nu}{\partial^2} \;.
\ee

Integrating out the field $\varphi$,
$$
{\cal Z}(J) \;=\; \int {\cal D}\chi^\dagger \,{\cal D} \chi \;
{\cal D}\Phi_\mu \,\delta(\partial \cdot \Phi)
\exp \left\{-\int d^3 x [\chi^\dagger(\not \! \partial + M) \chi \right.
$$
\be 
\left.
+ \, W(\Phi)- \frac{1}{2 g}\Phi_\mu \partial^2 {\cal P}_{\mu\nu}\Phi_\nu 
- \frac{1}{2 g} J_\mu {\cal P}_{\mu\nu} J_\nu
+ \frac{i}{g} J_\mu \epsilon_{\mu\nu\lambda}\partial_\nu \Phi_\lambda ] 
\right\} \;.
\ee
The term quadratic in the source $J_\mu$ can be, as in $1+1$ dimensions,
linearized by the introduction of an auxiliary field. In $2+1$
dimensions it has to be a vector field $\xi_\mu$, and the new expression
for ${\cal Z}$ becomes
$$
{\cal Z}(J) \;=\; \int {\cal D}\chi^\dagger \,{\cal D} \chi \;
{\cal D}\Phi_\mu {\cal D}\xi_\mu \,\delta(\partial \cdot \Phi)
$$
$$
\exp \left\{-\int d^3 x [\chi^\dagger(\not \! \partial + M) \chi 
+ W(\Phi) -\frac{1}{2 g} \Phi_\mu \partial^2 {\cal P}_{\mu\nu}\Phi_\nu 
+\frac{\lambda}{2}(\partial \cdot \xi)^2
\right.
$$
\be
\left.+\frac{\alpha}{2} \xi_\mu \partial^2 {\cal P}_{\mu\nu}\xi_\nu
+i J_\mu \epsilon_{\mu\nu\lambda} \partial_\nu (\frac{1}{g}\Phi_\lambda 
-\beta \xi_\lambda )]\right\} 
\ee
where the constants $\alpha$ and $\beta$ satisfy the relation: 
\be
\frac{\beta^2}{\alpha} = \frac{1}{g} \;.
\ee
and $\lambda \neq 0$ is arbitrary.
 
We then introduce two vector fields $\theta_\mu$ and $\theta_\mu'$,
defined in terms of the auxiliary fields $\Phi_\mu$ and $\xi$
such that: the transverse part of $\theta_\mu$ is uniquely determined by 
the requirement that it should be the one that couples to the external current,
and for the longitudinal part we choose it to be equal to the one of
the field $\xi$, i.e.,
\be
\theta_\mu \;=\; \frac{1}{g} {\cal P}_{\mu\nu} \Phi_\nu - 
\beta \xi_\mu \;,
\label{dfth}
\ee 
and $\theta_\mu'$ is instead defined in terms of $\Phi_\mu$ and
$\xi$ in such a way that the new action for the fields $\theta_\mu$ and 
$\theta_\mu'$ contains no mixing term.
There are many consistent ways of achieve this, however giving the same
bosonic action for the relevant field $\theta$,
but the simplest we found is 
\be
\theta_\mu' \;=\; \Phi_\mu \,-\,
[\frac{\partial^2}{C_+}\,{\cal P}_+
\,+\,\frac{\partial^2}{C_-} {\cal P}_-]_{\mu\nu} \theta_\nu 
\label{pet1}
\ee
with $\theta_\mu$, of course, as given in (\ref{dfth}). Note that,
by virtue of (\ref{pet1}), the longitudinal part of $\theta_\mu'$ 
is equal to the one of $\Phi$,
\be
\partial \cdot \Phi\;=\;\partial \cdot \theta' 
\ee 
thus the functional delta function of $\Phi$ becomes the gauge fixing
for $\theta'$.

The relevant bosonic field will be $\theta_\mu$, since by virtue of
the previous definitions $\theta_\mu'$ is decoupled and will
consequently be ignored. 
The resulting generating functional for 
$\theta_\mu$ and the decoupled fermions is
\newpage
$$
{\cal Z}(J) \;=\; \int {\cal D}\chi^\dagger \,{\cal D} \chi \;
{\cal D}\theta_\mu\, 
\exp \left\{-\int d^3 x [\chi^\dagger(\not \! \partial + M) \chi 
+\frac{\lambda}{2}(\partial \cdot \theta)^2 \right.
$$
\be
\left.+\frac{1}{2} \theta_\mu
[\frac{\partial^2}{C_+}(g C_+ - \partial^2) \,{\cal P}_+
\,+\,\frac{\partial^2}{C_-}(g C_- - \partial^2){\cal P}_-]_{\mu\nu}
\theta_\nu 
+ i J_\mu \epsilon_{\mu\nu\lambda}\partial_\nu \theta_\lambda] 
\right\} \;.
\ee
We can also write it more explicitly, in the notation of ref.~\cite{BFO}:
$$
{\cal Z}(J) \;=\; \int {\cal D}\chi^\dagger \,{\cal D} \chi \;
{\cal D}\theta_\mu\,
\exp \left\{-\int d^3 x [\chi^\dagger(\not \! \partial + M) \chi 
+\frac{\lambda}{2}(\partial \cdot \theta)^2 \right.
$$
\be
\left.-\frac{1}{4} F_{\mu\nu}(\theta) C_1  F_{\mu\nu}(\theta)
+\frac{i}{2}  \theta_\mu C_2 \epsilon_{\mu\nu\lambda} 
\partial_\nu \theta_\lambda - \frac{g}{4} F_{\mu \nu }^2(\theta ) 
+ i J_\mu \epsilon_{\mu\nu\lambda}\partial_\nu \theta_\lambda] 
\right\} 
\label{resultat}
\ee
where
\be
C_1\;=\;\frac{F}{(-\partial^2)F^2 + G^2} \;\;\;
C_2\;=\; \frac{G}{(-\partial^2)F^2 + G^2} \;.
\ee
and $F=F(-\partial^2)$, $G=G(-\partial^2)$ are given in terms of 
their Fourier transforms ${\tilde F}$ and ${\tilde G}$ by
\be
{\tilde F} (k^2) \;=\; \frac{\mid m \mid}{4 \pi k^2} \,
\left[ 1 - \displaystyle{\frac{1 \,-\,\displaystyle{\frac{k^2}{4 m^2}}}{(
\displaystyle{\frac{k^2}{4 m^2}})^{\frac{1}{2}}}} \, \arcsin(1\,+
\, \frac{4 m^2}{k^2})^{-\frac{1}{2}} \right] \;,
\ee
and
\be
{\tilde G} (k^2) \;=\; \frac{q}{4 \pi} \,+\, \frac{m}{2 \pi \mid k \mid}
\, \arcsin (1 \, + \, \frac{4 m^2}{k^2} )^{- \frac{1}{2}} \;,
\ee
where $q$ is an arbitrary integer.

Since $\chi $ and $\chi^\dagger$ are decoupled from the relevant 
bosonic field $\theta $, one recovers from (\ref{resultat}) the 
bosonization of the Thirring model in 3 dimensions, with a bosonic 
action given, in the quadratic approximation of $J(\Phi )$, by 
\be
S_{bos} \;=\; \int d^3 x [
-\frac{1}{4} F_{\mu\nu}(\theta) C_1  F_{\mu\nu}(\theta)
+\frac{i}{2}  \theta_\mu C_2 \epsilon_{\mu\nu\lambda} 
\partial_\nu \theta_\lambda 
- \frac{g}{4} F_{\mu \nu }^2(\theta )
+\frac{\lambda}{2}(\partial \cdot \theta)^2] 
\ee
and with the correspondence between the currents
\be
\Psi^\dagger(x)\gamma_\mu \Psi(x) \;\to \; i \epsilon_{\mu\nu\lambda} 
\partial_\nu \theta_\lambda(x) \;.
\label{ccc}
\ee

But the important new point is that, with our method, we are able to 
express the initial fermion fields in terms of the bosonic ones and 
free fermionic fields, which was not the case in the usual approach 
to bosonization in 3 dimensions, :
\be
\Psi = e^{-i\varphi } \left\{ 1-(\not\!\partial+M)^{-1}
\left(\not\!\partial 
[\gamma_\mu C_1 {\cal P}_{\mu\nu}\theta_\nu 
-i \gamma_\mu (-\partial^2)^{-1}C_2 
\epsilon_{\mu\nu\lambda}\partial_\nu \theta_\lambda
+\not\!\theta']\right) \right\}^{-\frac{1}{2}} \chi \;.
\ee
 
We end up this section by remarking that this redefinition of the
fermionic fields yields the proper expectation value for the current
(\ref{ccc}), as can be verified by an application of (\ref{pet2}) to this
case. 

\newpage
\section{The two ways of decoupling the fermions in $1+1$ dimensions}
We shall here compare the usual way of decoupling massless fermions in
$d=2$, which relies upon the anomalous Jacobian method, and ours, which
involves non-anomalous Jacobians, and moreover holds true in any number
of dimensions (in particular $d=2$). We perform the consistency check 
that both must lead to the same answer when applied to the two dimensional 
Abelian case. 

Let us begin by reviewing the anomalous Jacobian method. We start from the 
Euclidean generating functional
\be
{\cal Z}({\cal A})\;=\; \int {\cal D}{\bar \Psi} \, {\cal D}\Psi \;
\exp [ - \int d^2 x {\bar \Psi} (\not \! \partial + i e \not \!\!{\cal A})
\Psi ] \;.
\ee
The Dirac operator $\not \! \partial + i e \not \!\! {\cal A}$ may
be written as
\be
\not \! \partial + i e \not \!\! {\cal A}\;=\; \not \! \partial
+ i e \not \! \partial ({\not \! \partial}^{-1} \not \!\! {\cal A})
\ee
where we understand ${\not \! \partial}^{-1}$ as acting on ${\cal A}$ only, and
not on the fermionic field that may eventually appear on the right. Similarly, 
$\not \! \partial$ acts on $({\not \! \partial}^{-1}\not \!\!{\cal A})$.

Now, $({\not \! \partial}^{-1}\not \!\! {\cal A})$ {\em in $1+1$ dimensions\/}, 
leads in a natural way to the  decomposition of the vector field ${\cal A}_\mu$ 
into its longitudinal and transverse parts, since
\be
{\not \!\partial}^{-1} \not \!\! {\cal A} \;=\; \partial^{-2} \not \! \partial 
\not \!\!{\cal A} \;=\;\partial^{-2} 
(\partial \cdot {\cal A} + i \gamma_5 \epsilon_{\mu\nu}\partial_\mu 
{\cal A}_\nu )
\ee
or
\be
{\not \!\partial}^{-1} \not \!\! {\cal A} \;=\; \frac{1}{e} \not \! \partial 
(i \gamma_5 \Phi_1 + \Phi_2   )
\ee
where $\Phi_1= e \,\partial^{-2}\epsilon_{\mu\nu}\partial_\mu {\cal A}_\nu$ and 
$\Phi_2= e \, \partial^{-2}\partial \cdot {\cal A}$.

Then one may write the Dirac operator as
\be
\not \! \partial + i e \not \!\! {\cal A} \;=\; e^{-i \Phi_2 + i \gamma_5
\Phi_1} \not \! \partial e^{i \Phi_2 + i \gamma_5 \Phi_1}
\ee
and achieve decoupling by performing the transformation
\ba 
\Psi(x)&=&e^{-i\Phi_2(x)-i\gamma_5\Phi_1(x)} \chi(x)\nonumber\\
{\bar \Psi}(x)&=&{\bar \chi}(x)e^{i\Phi_2(x)-i\gamma_5\Phi_1(x)} \;.
\label{dft4}
\ea
After the transformation (\ref{dft4}) is performed, the action for the
new fermions is of course free
\be
S_F({\bar \Psi},\Psi;{\cal A}) \;=\; S_F({\bar \chi},\chi;0)
\label{free} 
\ee
and
there appears, due to the change of variables, an anomalous Jacobian
$J(\Phi_1)$
\be
J(\Phi_1) \;=\; \exp\left[-\frac{1}{2 \pi} \int d^2 x (\partial_\mu
\Phi_1)^2 \right]\;=\;
\exp [\frac{e^2}{4\pi} \int d^2 x F_{\mu\nu}
\frac{1}{\partial^2} F_{\mu\nu} ]
\;.
\ee
Thus, we have for ${\cal Z}({\cal A})$
\be
{\cal Z}({\cal A}) \;=\; \exp [\frac{e^2}{4\pi} \int d^2 x F_{\mu\nu}
\frac{1}{\partial^2} F_{\mu\nu} ]
\int {\cal D}{\bar \chi} \, {\cal D}\chi \;\exp [
\int d^2 x {\bar \chi} \not \!\partial \chi ] \;.
\label{aux2}
\ee

Let us now show that our way of decoupling yields to an identical
result. In Euclidean space, we must redefine the fields according to
\ba
\Psi (x) &=& (1 + ie {\not \! \partial}^{-1} \not \!\! {\cal A})^{-\frac{1}{2}}\,
\chi (x)\nonumber\\
\Psi^\dagger(x) &=& \chi^\dagger (x)(1 + 
ie \not \!\! {\cal A} {\not \! \partial}^{-1} )^{-\frac{1}{2}} \;,
\label{deft5}
\ea
after which the action becomes also free, as in the previous case.
The difference is that now we need to evaluate a (non anomalous)
Jacobian
\be
J({\cal A}) \;=\; \frac{\det (\not \!\! D)}{\det 
(\not \!\partial )} \;.
\label{aux1}
\ee
Again, due to the fact that we are considering the special case on
two spacetime dimensions, we may use the important result that
(\ref{aux1}) receives a non-vanishing contribution only from the
second order vacuum polarisation, which therefore is the exact
result for the logarithm of the Jacobian,
\be
J({\cal A}) \;=\; \exp [\frac{e^2}{4\pi} \int d^2 x F_{\mu\nu}
\frac{1}{\partial^2} F_{\mu\nu} ] \;,
\ee
and we have
\be
{\cal Z}({\cal A}) \;=\; \exp [\frac{e^2}{4\pi} \int d^2 x F_{\mu\nu}
\frac{1}{\partial^2} F_{\mu\nu} ]
\int {\cal D}\chi^\dagger \, {\cal D}\chi \;\exp [
\int d^2 x \chi^\dagger \not \!\partial \chi ] \;,
\ee
to be compared with (\ref{aux2}).
\newpage
\section*{Acknowledgements}
C.D.F. acknowledges the members of the Laboratoire  d'Annecy-le-Vieux
de Physique Th\'eorique LAPTH, where this work was done, for their
kind hospitality. 

\end{document}